\documentclass[aps,amsmath,onecolumn,amssymb,nofootinbib,floatfix,showpacs]{revtex4}
\usepackage{graphicx}
\usepackage{amsmath}
\usepackage{amssymb}
\usepackage{latexsym}
\usepackage{dcolumn}
\usepackage{bm}
\usepackage[dvipdfmx]{hyperref}
\usepackage{url}
\tighten

\begin{document}
\title{The Friedberg-Lee model at finite temperature and density }
\author{Hong Mao$^{1,3}$}
\email {maohong@hznu.edu.cn}
\author{Minjie Yao$^{2}$}
\author{Wei-Qin Zhao$^{3}$}
\address{1. Department of Physics, Hangzhou Normal University,
Hangzhou 310036, China  \\
2. Department of Mathematics, Hangzhou Normal University, Hangzhou
310036, China \\3. CCAST(World Laboratory), P.O. Box 8730, Beijing
100080, China}


\begin{abstract}
The Friedberg-Lee model is studied at finite temperature and
density. By using the finite temperature field theory, the effective
potential of the Friedberg-Lee model and the bag constant $B(T)$ and
$B(T,\mu)$ have been calculated at different temperatures and
densities. It is shown that there is a critical temperature
$T_{C}\simeq 106.6 \mathrm{MeV}$ when $\mu=0 \mathrm{MeV}$ and a
critical chemical potential $\mu \simeq 223.1 \mathrm{MeV}$ for
fixing the temperature at $T=50 \mathrm{MeV}$. We also calculate the
soliton solutions of the Friedberg-Lee model at finite temperature
and density. It turns out that when $T\leq T_{C}$ (or $\mu \leq
\mu_C$), there is a bag constant $B(T)$ (or $B(T,\mu)$) and the
soliton solutions are stable. However, when $T>T_{C}$ (or
$\mu>\mu_C$) the bag constant $B(T)=0 \mathrm{MeV}$ (or $B(T,\mu)=0
\mathrm{MeV}$) and there is no soliton solution anymore, therefore,
the confinement of quarks disappears quickly.
\end{abstract}


\maketitle

\section{Introduction}

It is widely believed that at sufficiently high temperatures and
densities there is a deconfinement phase transition between normal
nuclear matter and quark-gluon plasma (QGP), where quarks and gluons
are no longer confined in
hadrons\cite{Philipsen:2007rj}\cite{Petreczky:2007js}. The
theoretical and experimental investigation of QGP is one of the most
challenging problems in high energy physics. The main goal of heavy
ion experiments is to create such form of matter and study its
properties.

Due to the property of confinement and asymptotic freedom, QCD can
not yet be used to study low-energy nuclear physics. The challenge
to nuclear physicists is to find models which can bridge the gap
between the fundamental theory and our wealth of knowledge about low
energy phenomenology. Some of these models have been proved to be
successful in reproducing different properties of hadrons, nuclear
matter and quark matter
\cite{Serot:1997xg,Tomas:1984,Farhi:1984qu,Madsen:1989pg,Greiner:1987tg,Fowler:1981rp},
among which Friedberg-Lee model\cite{Friedberg:1976eg}, also
referred to as nontopological soliton model\cite{Goldflam:1981tg},
has been widely discussed in the past decades, see for instance
Refs.\cite{Lee:1991ax}\cite{Wilets:1989}\cite{Birse:1991cx}. The
model consists of quark fields interacting with a phenomenological
scalar field $\sigma$. The $\sigma$ field is introduced to describe
the complicated nonperturbative features of the QCD vacuum. It shows
an intuitive mechanism for the deconfinement phase transition. In
physical vacuum state, the physical value of $\sigma$ is large and
quark mass is more than $1 \mathrm{GeV}$ which makes it
energetically unfavorable for the quark to exist freely, so that the
effective heavy quarks have to be confined in hadron bags. But with
the temperature and/or density of the system increasing, the
physical value of $\sigma$ decreases and in turn the effective quark
mass drops down, the thermodynamic motion leads to a deconfinement
of the effective light quarks.

The properties of Friedberg-Lee model, mostly in the mean field
approximation, are widely studied by Wilets and his
co-workers\cite{Goldflam:1981tg}\cite{Wilets:1989}. The model has
been very successful in describing the static properties of isolated
hadrons. Moreover, during the past several years, Friedberg-Lee
model was extended to finite temperature and density to study
deconfinement phase
transitions\cite{Reinhardt:1985nq}\cite{Gao:1992zd}\cite{Li:1987wb}\cite{Wang:1989ex}\cite{Yang:2003pz}.
In the framework of finite temperature field theory, most of those
studies focus on the effective potential of the Friedberg-Lee model
at finite temperature and density, but seldom considered the value
of the soliton configurations at high temperature and density. In
Our previous work\cite{Mao:2006zp} the improved quark mass
density-dependent model(IQMDD) was studied at finite temperature. By
using the finite temperature field theory, the effective potential
of the IQMDD model and the bag constant $B(T)$ have been calculated
at different temperatures, and it is shown that there is a critical
temperature at $T\simeq 110 MeV$ when the deconfinement phase
transitions take place. In the meantime, the soliton solutions at
different temperatures are obtained. The Lagrangian of the
Friedberg-Lee model we use is almost identical to that of IQMDD
model which was studied in our previous work, especially their
potentials take the same forms, but there are two basic differences
between the Friedberg-Lee model and IQMDD model suggested first in
Ref.\cite{Wu:2005ty}. First, for the the Friedberg-Lee model, inside
the bag (a region of the metastable vacuum), the quarks have zero
rest mass and only kinetic energy, on the contrary, according to the
IQMDD model, the masses of $u,d$ quarks are given by
$m_q=\frac{B}{3n_B} (q=u, d,\bar{u},\bar{d})$, $B$ is the vacuum
energy density inside the Bag, and $n_B$ is the baryon number
density. Second, in the Friedberg-Lee model, the $\sigma$ field is a
phenomenological scalar field which is introduced to describe the
complicated nonperturbative features of the QCD vacuum, but for the
IQMDD model, the $\sigma$ field is the meson field, and the coupling
of quarks and the meson field $\sigma$ is introduced to improve the
QMDD model\cite{Fowler:1981rp}
 by including the quark-quark interactions. So that the parameters of the Friedberg-Lee model are
 significantly different from the ones in the IQMDD model, but the
 discussions adopted in the previous one can still be applied.

In the resent years, the research of the the deconfinement phase
transition at finite temperature and density has received a great
deal of attention, then it is interesting to extend the
Friedberg-Lee model to the finite temperature and density to study
the detailed behaviors of the soliton solutions at different
temperatures and densities. Since there are lots of works on the
effective potential at finite temperature, in present paper, we
especially focus on the study of the value of the soliton
configurations at different temperatures and densities, in doing so
we can give more detailed information of the the deconfinement phase
transition.

The organization of this paper is as follows. In the following
section we review the Friedberg-Lee model and our selected
parameters. In Sec.III, we give detailed calculation of the
effective potential of the Friedberg-Lee model and the bag constant
as functions of temperature and density. The soliton solutions of
the Friedberg-Lee model at different temperatures and densities are
presented in Sec.IV, while in the last section we present our
summary and discussion.

\section{The Friedberg-Lee model}
The Lagrangian of the Friedberg-Lee model has the
form\cite{Friedberg:1976eg}
\begin{eqnarray}\label{lagrangian}
\mathcal{L}=\overline{\psi}(i\eth -g
\sigma)\psi+\frac{1}{2}\partial_{\mu}\sigma
\partial^{\mu}\sigma -U(\sigma),
\end{eqnarray}
which describes the interaction of the spin-$\frac{1}{2}$ quark
fields $\psi$ and the phenomenological scalar field $\sigma$ with
the coupling constant $g$, $U(\sigma)$ is a potential describing the
nonlinear self-interactions of the $\sigma$ field. We will discuss
only u and d quark in this paper. The potential for the $\sigma$
field is chosen as
\begin{eqnarray}
U(\sigma)=\frac{a}{2!}\sigma^2+\frac{b}{3!}\sigma^3+\frac{c}{4!}\sigma^4+B,
\end{eqnarray}
\begin{equation} \label{eq01}
b^2 > 3ac
\end{equation}
The model parameters $a$, $b$ and $c$ are fixed so that $U(\sigma)$
has a local minimum at $\sigma_0=0$,  and a absolute minimum at a
large value of the $\sigma$ field
\begin{eqnarray}
\sigma_v=\frac{3|b|}{2c}\left[1+\left[1-\frac{8ac}{3b^2}\right]^{\frac
1 2}\right].
\end{eqnarray}
The condition (\ref{eq01}) ensures that the absolute minimum of
$U(\sigma)$ is at $\sigma = \sigma_{v} \ne 0$. $\sigma_0$ represents
a metastable vacuum where the condensates vanishes, $\sigma_v$
corresponds to the physical or nonperturbative vacuum. The
difference in the potential of the two vacuum states is the bag
constant $B$. If we take $U(\sigma_{\nu})=0$, the bag constant $B$
can be expressed as
\begin{eqnarray}
-B=\frac{a}{2!}\sigma^2_v+\frac{b}{3!}\sigma^3_v+\frac{c}{4!}\sigma^4_v.
\end{eqnarray}

The Euler-Lagrange euqations corresponding to (\ref{lagrangian}) are
given by
\begin{eqnarray}\label{euler-eq1}
(i\eth-g\sigma)\psi &=& 0, \\ \square
\sigma+\frac{dU}{d\sigma}+g\overline{\psi}\psi &=& 0.
\label{euler-eq2}
\end{eqnarray}
In the mean-field approximation (MFA), the scalar field $\sigma$ is
taken as a time-independent classical c-number field, and we only
consider a fixed occupation number of valence quarks (3 quarks for
nucleons, and a quark-antiquark pair for mesons). Quantum
fluctuation of the bosons and effects of the quark Dirac sea are
thus to be neglected.

In the following, by solving the Euler-Lagrange euqations, the
numerical calculation indicates that the $\sigma$ field has a
soliton solution which is of a spherical cavity-like structure, the
metastable vacuum is inside the cavity while the physical vacuum is
outside, and the quarks are confined inside the cavity, however,
with increase of temperature or density, the solutions of the
equation of motions will become a damping oscillation, i.e. the
soliton solutions are melted away and the deconfinement phase
transitions take place.

In the spherical case, the $\sigma$ field is spherically symmetric,
and valence quarks are in the lowest s-wave level. Then the scalar
field $\sigma$ and the Dirac equation functions can be written as
\begin{eqnarray}
\sigma(\mathbf{r},t) &=& \sigma(r), \\
\psi(\mathbf{r},t) &=& e^{-i\epsilon t}\sum_{i}\varphi_i,
\end{eqnarray}
where the quark Dirac spinors have the form
\begin{eqnarray} \label{spinor}
\varphi=\left(\begin{array}{c} u(r) \\
i \vec{\sigma} \cdot\mathbf{\hat{ r}}v(r)
\end{array}\right)\chi.
\end{eqnarray}
By using Eqs.(\ref{euler-eq1})-(\ref{spinor}), we obtain
\begin{eqnarray}
\frac{du(r)}{dr}=-(\epsilon+g \sigma(r))v(r), \label{equation1}\\
\frac{dv(r)}{dr}=-\frac{2}{r}v(r)+(\epsilon-g \sigma(r))u(r), \label{equation2} \\
\frac{d^2
\sigma(r)}{dr^2}+\frac{2}{r}\frac{d\sigma(r)}{dr}-\frac{dU}{d\sigma}=Ng(u^2(r)-v^2(r)).
\label{equation3}
\end{eqnarray}
The quark functions should satisfy the normalization condition
\begin{eqnarray}
4\pi \int r^2 (u^2(r)+v^2(r))dr=1.
\end{eqnarray}
The number of quarks is $N=3$ for baryons and $N=2$ for mesons. In
the following, our discussions are constrained in the case of $N=3$.
These equations are subject to the boundary conditions which follow
from the requirement of finite energy:
\begin{eqnarray}
v(0)=0,     \frac{d\sigma(0)}{dr}=0,\\
u(\infty)=0,          \sigma(\infty)=\sigma_v.
\end{eqnarray}

If we consider $N$ quarks in the lowest mode with energy $\epsilon$,
the total energy of the system is
\begin{eqnarray}\label{energy}
E=N \epsilon+4\pi\int r^2 \left[ \frac{1}{2}
\left(\frac{d\sigma}{dr}\right)^2+U(\sigma) \right]dr.
\end{eqnarray}

The model has four adjustable parameters $g$, $a$, $b$, $c$ which
can be chosen to fit various baryon properites, such as masses,
charge radii and magnetic moments. Let's set $R$, $\mu_p$ and
$g_A/g_V$ be the proton charge radius, the proton magnetic moment
and the ratio of axial-vector to vector coupling respectively, then
they are given by
\begin{equation}\label{radius}
R^2=4\pi\int^\infty_0 r^4(u^2(r)+v^2(r))dr,
\end{equation}
\begin{equation}
\mu_p=\frac{8\pi}{3}\int^\infty_0 r^3 u(r)v(r)dr,
\end{equation}
\begin{equation}
\frac{g_A}{g_V}=\frac{20\pi}{3}\int^\infty_0
r^2(u^2(r)-\frac{1}{3}v^2(r))dr.
\end{equation}
As in any simple quark model, the neutron magnetic moment is
$-\frac{2}{3}$ that of the proton, and its charge radius is zero.
Once the solutions to the above equations are obtained, one can
calculate these physical quantities pertaining to the three-quark
system, which have been measured experimentally. In
Ref.\cite{Goldflam:1981tg,Li:1987wb,Gao:1992zd}, a wide range of
parameters have been used to calculate the quantities above.
Hereafter we take one set of parameters $a=17.70 fm^{-2}$,
$b=-1457.4 fm^{-1}$, $c=20000$ and $g=12.16$ to study the
temperature and density dependence of the soliton solution. It has
been proved in Ref.\cite{Goldflam:1981tg,Li:1987wb,Gao:1992zd} that
this set of parameters can describe the properties of nucleon at
zero temperature successfully.

\section{The one-loop effective potential}
A convenient framework of studying phase transitions is the thermal
field theory. Within this framework, the finite temperature
effective potential is an important and useful theoretical tool. In
this section, in order to investigate the temperature and the
chemical potential dependence of the Friedberg-Lee model, we
summarize the relevant results of Dolan and
Jackiw\cite{Dolan:1973qd}, since the model we use is almost
identical to the model studied by those authors.

The finite temperature partition function $Z(J;\beta)$ is given by
\begin{eqnarray}
Z(J;\beta)=\frac{\mathrm{Tr}[e^{-\beta H} \mathrm{T}
\mathrm{exp}(iJ\cdot\sigma) ]}{\mathrm{Tr}[e^{-\beta H}]},
\end{eqnarray}
where $H$ is the Hamiltonian obtained from the
Lagrangian(\ref{lagrangian}) and $\beta$ is the inverse of the
temperature ($\beta=T^{-1}$). $\mathrm{T}
\mathrm{exp}(iJ\cdot\sigma)$ denotes the time-ordered exponential of
\begin{eqnarray}
iJ\cdot\sigma\equiv i\int d^4x J(x)\sigma (x),
\end{eqnarray}
where, in the imaginary-time formalism, $x^0$ runs from $0$ to $-i
\beta$. The $\sigma$ field is required to satisfy periodic boundary
conditions in time; the $\psi$ fields are antiperiodic in time. The
generating function for the connected Green functions is given in
terms of $Z(J;\beta)$ by
\begin{eqnarray}
W(J;\beta)=-i \mathrm{ln} Z(J;\beta).
\end{eqnarray}
The effective action is obtained via a Legendre transformation of
$W(J;\beta)$,
\begin{eqnarray}
\Gamma (\overline{\sigma};\beta)=W(J;\beta)-\int d^4x
\overline{\sigma}J(x),
\end{eqnarray}
where the average value of the $\sigma$ field is
\begin{eqnarray}
\overline{\sigma}(x)=\frac{\delta W(J;\beta)}{\delta J(x)}.
\end{eqnarray}
Here we study a uniform system in which $\overline{\sigma}$ does not
depend on the space-time coordinates. The generalized effective
potential for such a system is defined by
\begin{eqnarray}
V(\overline{\sigma};\beta)=-\Omega^{-1}
\Gamma(\overline{\sigma};\beta),
\end{eqnarray}
where an infinite volume factor $\Omega$ arises from space-time
integrations and it is customary to introduce the generalized
effective potential $V(\overline{\sigma};\beta)$. In the
thermodynamic language, the thermal effective potential has the
meaning of the free energy density and it is related to the
thermodynamic pressure through the equation
\begin{eqnarray}
p=-V(\overline{\sigma};\beta).
\end{eqnarray}

The effective potential can be calculated to one-loop order by using
the methods of Dolan and Jackiw\cite{Dolan:1973qd}. However, it is
pointed out in Ref\cite{Friedberg:1976eg, Lee:1981mf} that, as an
approximation, all $\sigma$ quantum loop diagrams may be ignored due
to the fact that $\sigma$ is only a phenomenological field
describing the long-range collective effects of QCD, and its
short-wave components do not exist in reality. Therefore for the
rest of this discussion we shall ignore quantum corrections and
concentrate on those induced by finite temperature and density
effects.

In order to study the temperature and density dependence of the bag
constant and the features of the deconfinement phase transitions, we
first calculate the effective potential at finite temperature and
density. Using the techniques of Dolan and
Jackiw\cite{Dolan:1973qd}, the one-loop contribution to the
effective potential of the Friedberg-Lee model at finite temperature
and density is of the form
\begin{eqnarray}\label{potential0}
V(\sigma;\beta,\mu)=U(\sigma)+V_B(\sigma;\beta)+V_F(\sigma;\beta,\mu),
\end{eqnarray}
where $U(\sigma)$ is the classical potential of the Lagrangian.
$V_B(\sigma;\beta)$ is the finite temperature contributions from
boson one-loop diagrams, and $V_F(\sigma;\beta,\mu)$ is the finite
temperature and density contributions from fermion one-loop
diagrams\cite{Mao:2006zp}\cite{Dolan:1973qd}. For simplicity, we
have substituted $\sigma(T)$ for $\bar{\sigma}(T)$. These contribute
the following terms in the potential\cite{Dolan:1973qd}
\begin{eqnarray}\label{potential1}
V_B(\sigma;\beta)=\frac{1}{2\pi^2 \beta^4} \int^{\infty}_0 dx x^2
\mathrm{ln} \left( 1-e^{-\sqrt{x^2+\beta^2 m_{\sigma}^2}} \right),
\end{eqnarray}
\begin{eqnarray}\label{potential2}
V_F(\sigma;\beta,\mu)=-12\sum_n \frac{1}{2\pi^2 \beta^4}
\int^{\infty}_0 dx x^2 \mathrm{ln} \left( 1+e^{-(\sqrt{x^2+\beta^2
m_{q}^2}-\beta\mu)} \right),
\end{eqnarray}
where the minus sign is the consequence of Fermi-Dirac statistics,
$m_{\sigma}$ and $m_q$ are the effective masses of the scalar field
$\sigma$ and the quark field, respectively:
\begin{eqnarray}
m_q &=& g \sigma(T), \label{massq}\\
m^2_{\sigma}&=& a+b \sigma(T)+\frac{c}{2} \sigma^2(T). \label{masss}
\end{eqnarray}
We fix $m^2_{\sigma}$ by taking its value at the physical vacuum
state\cite{Gao:1992zd}. In this paper, the bag constant $B(T,\mu)$
is defined as the difference between the vacua of the effective
potential inside and outside the solion bag. This means that for
$T\leq T_{C}$ (or $\mu \leq \mu_C$), $B(T,\mu)$ is the difference
between the effective potential values at the perturbative vacuum
state and values at the physical vacuum state
\begin{eqnarray}\label{bag}
B(T,\mu)=V(\sigma_0;\beta,\mu)-V(\sigma_v;\beta,\mu).
\end{eqnarray}
For $T > T_{C}$ (or $\mu > \mu_C$), the bag constant is zero due to
the fact that the vacua inside and outside the soliton bag are
equal. This will be analyzed in detail in next section. From
Eqs.(\ref{potential0}), (\ref{massq}) and (\ref{masss}), we can
numerically solve the effective potential $V(\sigma;\beta,\mu)$ for
different temperatures and densities.

\begin{figure}
\includegraphics[scale=0.36]{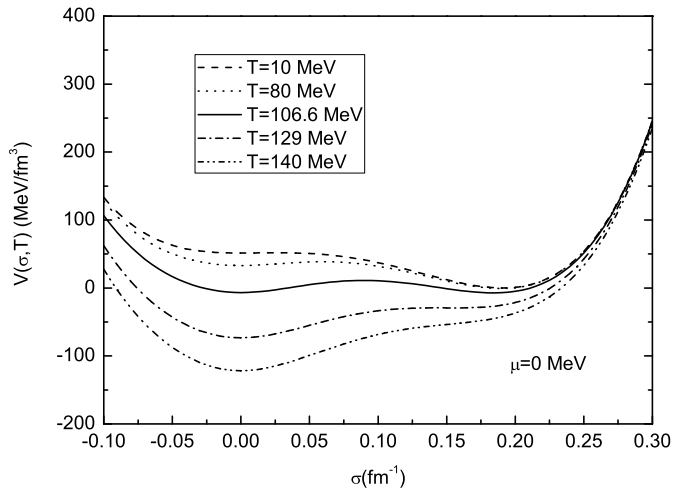}
\caption{\label{Fig:Fig3} The one-loop Effective potential
$V(\sigma;\beta)$ as a function of $\sigma$ at $T=10 \mathrm{MeV}$,
$T=80 \mathrm{MeV}$,$T=106.6 \mathrm{MeV}$, $T=129 \mathrm{MeV}$ and
$T=140 \mathrm{MeV}$ when the chemical potential $\mu=0$. The
critical temperatures is set at $T_{C}\simeq 106.6 \mathrm{MeV}$
when two minima are euqal.}
\end{figure}

When the chemical potential is zero, the effective potential at
different temperatures is illustrated in Fig.\ref{Fig:Fig3}. It can
be seen from Fig.\ref{Fig:Fig3} that there exist two particular
temperatures. One is that the effective potential exhibits two
degenerate minima at $T_{C}\simeq 106.6 \mathrm{MeV}$ which is
defined as the critical temperature, the other is that the second
minimum of the potential at $\sigma\simeq \sigma_v$ disappears at a
higher temperature $T\simeq 129 \mathrm{MeV}$.  For low temperatures
the absolute minimum of $V(\sigma;\beta)$ lies close to $\sigma_v$,
and there is another minimum at $\sigma_0$. The physical vacuum
state at $\sigma_v$ is stable, and correspondingly quarks are in
confinement. As the temperature increases the second minimum of the
potential at $\sigma_0$ decreases relative to the first one. At the
critical temperature $T_{C}\simeq 106.6 \mathrm{MeV}$, the
potentials at the two minima are equal. The physical vacuum becomes
unstable and soliton solutions are going to disappear.

\begin{figure}
\includegraphics[scale=0.36]{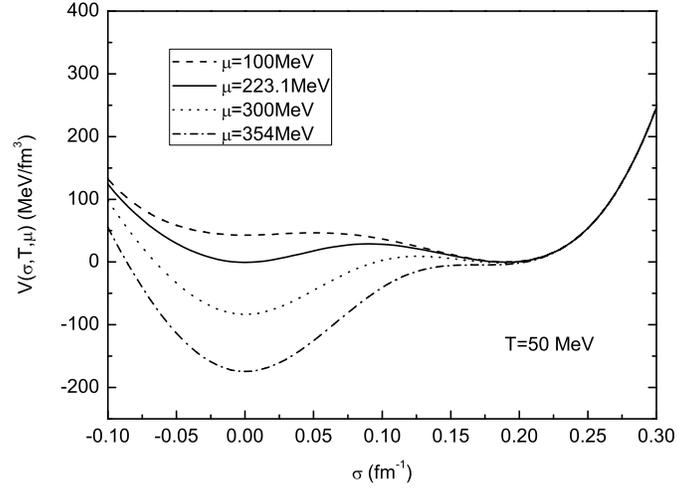}
\caption{\label{Fig:Fig4} The temperature and density dependence of
the one-loop effective potential $V(\sigma;\beta,\mu)$ when fixing
temperature $T$ at $50$ MeV. At low chemical potential, there are
two minima and one of which is going to disappear when the chemical
potential reaching a high value $\mu=354 \mathrm{MeV}$. The critical
chemical potential is set at $\mu_{C}\simeq 223.1 \mathrm{MeV}$ when
two minima are euqal.}
\end{figure}

When fixing temperature $T=50 \mathrm{MeV}$, in Fig.\ref{Fig:Fig4}
we plot the one-loop Effective potential $V(\sigma;\beta,\mu)$ as a
function of $\sigma$ at difference chemical potentials $\mu=100
\mathrm{MeV}$, $\mu=223.1 \mathrm{MeV}$, $\mu=300 \mathrm{MeV}$ and
$\mu=354 \mathrm{MeV}$ . From Fig.\ref{Fig:Fig4} , there are two
specific chemical potentials, one corresponds to the critical
chemical potential $\mu_{C}\simeq 223.1 \mathrm{MeV}$ where the
effective potentials at the two vacuum states are equal, the other
is a higher chemical potential $\mu \simeq 354 \mathrm{MeV}$ at
which the chemical potential increases and the second minimum of the
potential at $\sigma\simeq \sigma_v$ disappears. For low chemical
potentials the absolute minimum of $V(\sigma;\beta,\mu)$ lies close
to $\sigma_v$, and there is another relative minimum at $\sigma_0$.
The physical vacuum state at $\sigma_v$ is stable and
correspondingly quarks are in confinement. As the chemical potential
increases the second minimum of the potential at $\sigma_0$
decreases relative to the first one. At the critical chemical
potential $\mu_{C}\simeq 223.1 \mathrm{MeV}$, the potentials at the
two minima are equal. When the chemical potential increases further,
the physical vacuum becomes unstable and the soliton solutions tend
to melt away.

\begin{figure}
\includegraphics[scale=0.36]{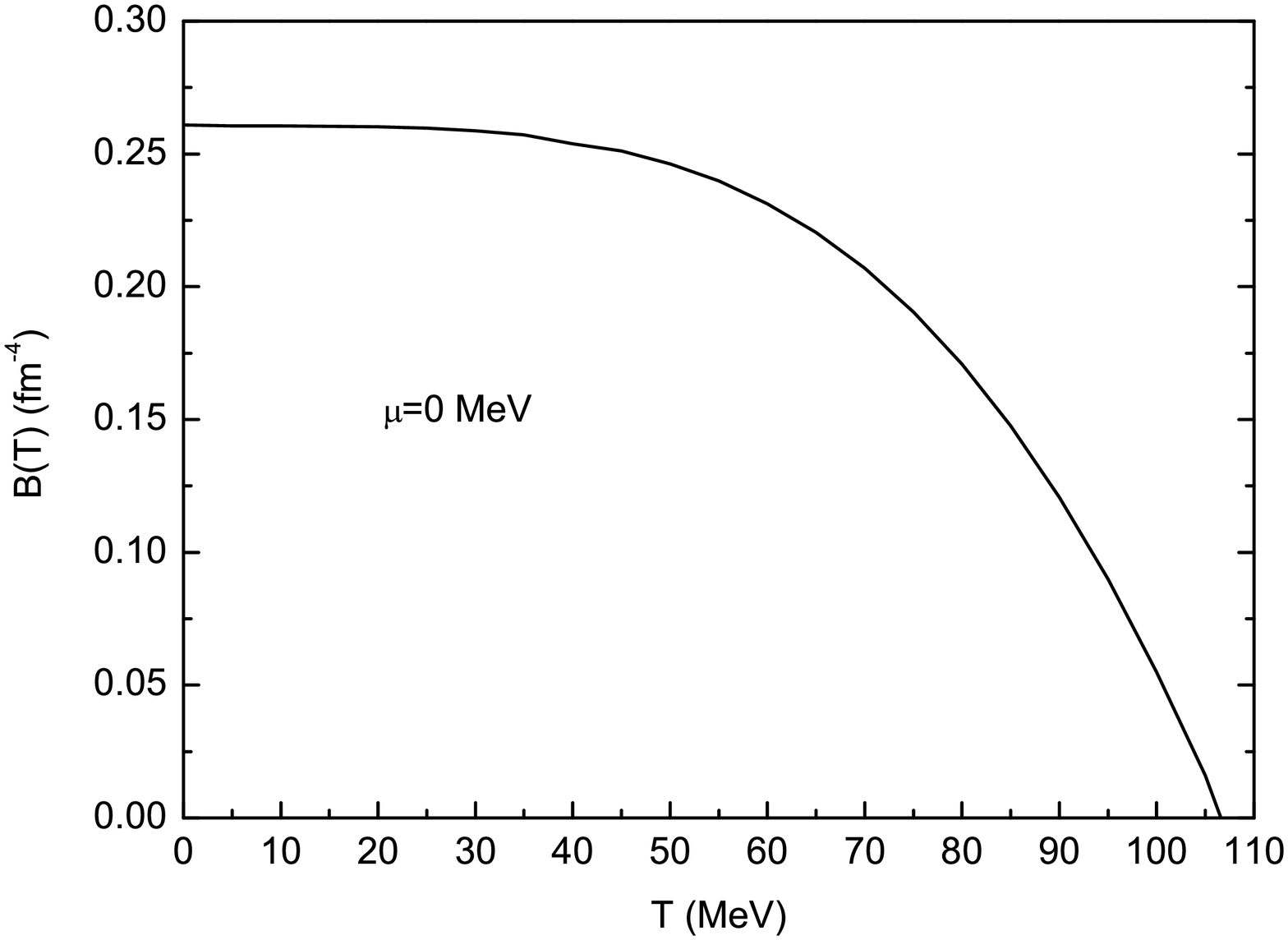}
\caption{\label{Fig:Fig5} The bag constant $B(T)$ as functions of
$T$ when the chemical potential is zero.}
\end{figure}

\begin{figure}
\includegraphics[scale=0.36]{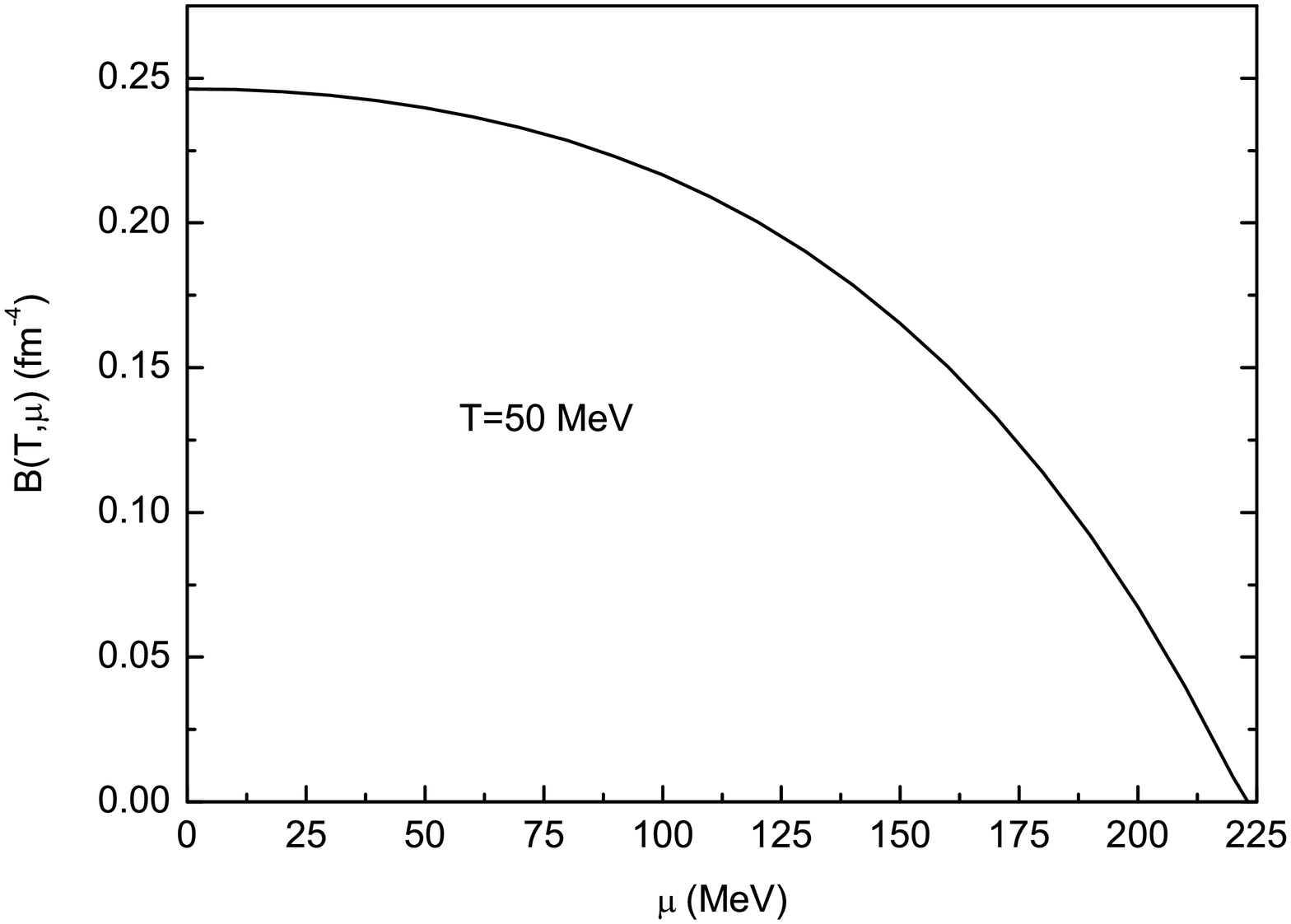}
\caption{\label{Fig:Fig6} The bag constant $B(T,\mu)$ as functions
of $\mu$ when the temperature is fixed at $T=50 \mathrm{MeV}$.}
\end{figure}

From the above effective potential, it is convenient to investigate
the temperature and the chemical potential dependence of the bag
constant. For the case of finite temperature and zero density, the
bag constant $B(T)$ in the the Friedberg-Lee model is defined as the
difference between the vacua of the effective potential inside and
outside the solion bag. This means that for $T\leq T_C$, the bag
constant is defined as the difference between the effective
potential at the perturbative vacuum state and the effective
potential at the physical vacuum state,
\begin{eqnarray}\label{bag1}
B(T)=V(\sigma_0;\beta)-V(\sigma_v;\beta).
\end{eqnarray}
For $T > T_{C}$, $B(T)=0$. From Eq.(\ref{bag1}) we illustrate the
temperature dependence of the bag constant $B(T)$ in
Fig\ref{Fig:Fig5}. Since $B(T)=0$ when $T$ is lager than $T_{C}$,
there is no bag constant to provide the dynamical mechanism to
confine the quarks in the bag. Moreover, there is no soliton-like
solution in the model when the temperature is above $T_C$ due to the
effective potential and vacuum structure, there only exists the
damping oscillation solution which can not produce a mechanism to
confine the quarks in a small region, we will investigate the
damping oscillation next section in details. Therefore, as the
temperature is above $T_{C}$, the confinement of the quarks is
removed completely. Similarly, for the case of fixing temperature at
$T = 50 \mathrm{MeV}$, the bag constant at different densities is
defined as
\begin{eqnarray}\label{bag2}
B(T,\mu)=V(\sigma_0;\beta,\mu)-V(\sigma_v;\beta,\mu).
\end{eqnarray}
And for $\mu > \mu_{C}$, $B(T,\mu)=0$. The result is shown in
Fig.\ref{Fig:Fig6}. In this case, we can see the deconfinement phase
transitions begin to take place at the chemical potential $\mu_C$.

\begin{figure}
\includegraphics[scale=0.36]{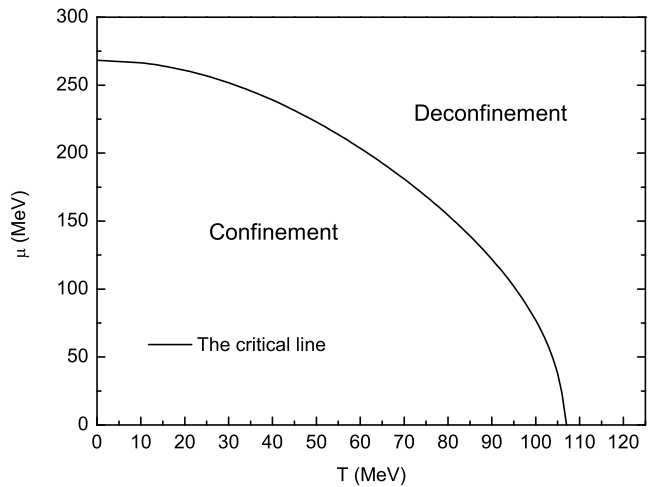}
\caption{\label{Fig:Fig13} The phase diagrams for the Friedberg-Lee
model in the ($T-\mu$) plane. Along this line the effective
potential at the two vacuum states are equal, the physical vacuum
becomes unstable and the soliton solutions tend to melt away.}
\end{figure}

In the end of this section, by using the similar discussion in
Ref.\cite{Scavenius:2000qd}, the phase diagram in the $T-\mu$ plane
calculated for the Friedberg-Lee model is shown in
Fig.\ref{Fig:Fig13}. The critical line corresponds to the phase
transition line between confined and deconfined phases. At the
critical $T$ and $\mu$ the two vacumm states, namely the
perturbative vacumm state and the physical vacumm state, appear in
the same energy. This is treated as the sign of the first order
phase transition\cite{Gao:1992zd}\cite{Wang:1989ex}
\cite{Scavenius:2000qd}. Our phase diagram shown in
Fig.\ref{Fig:Fig13}, based on the phenomenological Friedberg-Lee
model, predicts a first order deconfinement phase transition for the
full phase diagram since there are two equal minima in the effective
potential separated by the barrier. This result is different from
the predictions based on lattice gauge theory where more complicated
phase diagram on the $T$ and $\mu$ plane has been obtained, giving
more fruitful behaviors of the QCD
phase\cite{Kapusta:2006pm}\cite{Rischke:2003mt}. For example, at
finite temperature and zero chemical potential, the phase transition
could be of a second-order, moreover, it is predicted that there may
exist a critical point along the critical line, at lower baryon
density there might be a rapid crossover, while at higher baryon
density, the first order phase transition is obtained. Finally, we
would like to point out that the assumed phenomenological potential
in the Friedberg-Lee model should be the reason for these
differences in the behavior of phase diagram.

\section{The soliton solutions at finite temperature and density}

Field models such as those described above can be modified so as to
allow for the effect of a thermal background, with temperature $T$,
just by replacing the relevant classical potential function
$U(\sigma)$ by an appropriately modified temperature-dependent
effective potential
$V(\sigma;\beta)$\cite{Mao:2006zp}\cite{Holman:1992rv}\cite{Carter:2002te}.
In this work, in order to study the effects of the temperature and
density on the soliton solutions of the Friedberg-Lee model, the
classical potential function $U(\sigma)$ should be replaced by the
temperature and density dependent effective potential
$V(\sigma;\beta,\mu)$.

As in the zero temperature and zero chemical potential case, for
finite temperature and density soliton solutions will take the same
form as the one at zero temperature and zero chemical potential. But
the $\sigma$ field should be determined by the equation of motion:
\begin{eqnarray}\label{equation31}
\frac{d^2
\sigma(r)}{dr^2}+\frac{2}{r}\frac{d\sigma(r)}{dr}-\frac{dV(\sigma;\beta,\mu)}{d\sigma}=Ng(u^2(r)-v^2(r)),
\end{eqnarray}
where the temperature and density dependent effective potential
$V(\sigma;\beta,\mu)$ is defined in Eq.(\ref{potential0}). And the
quark functions should also satisfy the normalization condition
\begin{eqnarray}
4\pi \int r^2 (u^2(r)+v^2(r))dr=1
\end{eqnarray}

\begin{figure}
\includegraphics[scale=0.36]{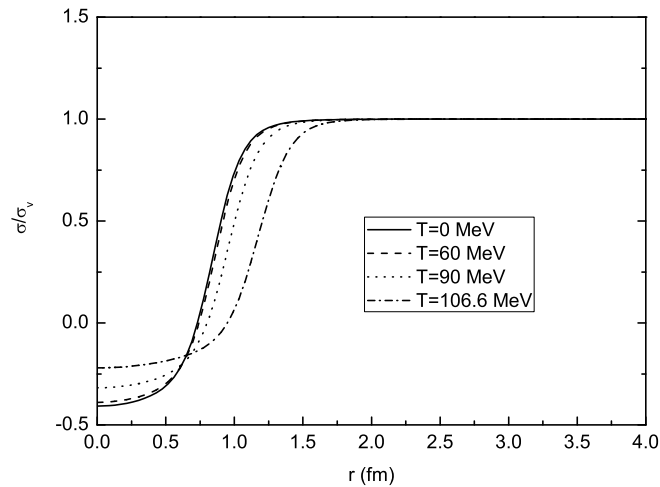}
\caption{\label{Fig:Fig7} The solutions for the set of coupled
nonlinear differential equations (\ref{equation1}),
(\ref{equation2}) and (\ref{equation31}) for different temperatures
when $T\leq T_{C}$ and $\mu=0 \mathrm{MeV}$.}
\end{figure}

\begin{figure}
\includegraphics[scale=0.36]{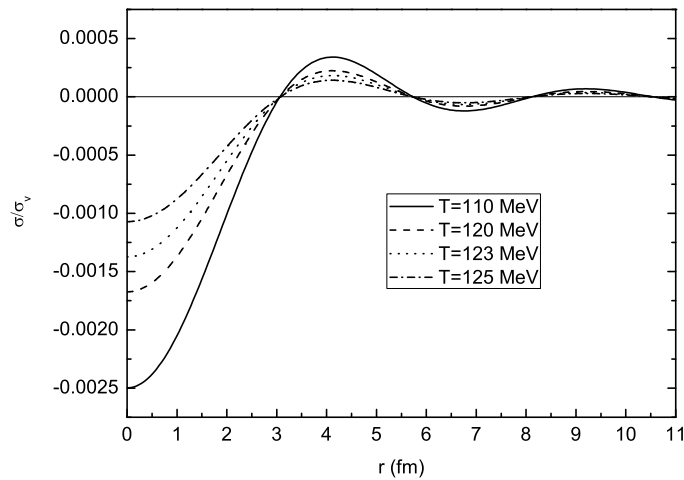}
\caption{\label{Fig:Fig11} The solutions for the set of coupled
nonlinear differential equations (\ref{equation1}),
(\ref{equation2}) and (\ref{equation31}) for different temperatures
when $T>T_{C}$ and $\mu=0 \mathrm{MeV}$.}
\end{figure}

The functions $\sigma(r)$, $u(r)$ and $v(r)$ also satisfy the
boundary conditions following from the requirement of finite energy:
\begin{eqnarray} \label{condition1}
v(0)&=& 0,   \frac{d\sigma(0)}{dr}=0, \label{condition1}\\
u(\infty)&=& 0.\label{condition2}
\end{eqnarray}
However, the situation is changed according to $\sigma(r)$ as
$r\rightarrow \infty$. When $T\leq T_{C}$ (or $\mu\leq \mu_{C}$), in
order to satisfy the requirement of finite energy of the solition
(or other topological defects), as $r\rightarrow \infty$,
 $\sigma(r)$ should be equal to $\sigma_v$, where the potential
$V(\sigma)$ has an absolute minimum. For $T> T_{C}$  (or $\mu>
\mu_{C}$), the physical vacuum becomes unstable, and the stable
vacuum  is the perturbative vacuum which is the absolute minimum of
the effective potential. Therefore, to satisfy the requirement of
finite energy of the solition at $T> T_{C}$ (or $\mu> \mu_{C}$), we
should take the asymptotic value (vacuum values) $\sigma \rightarrow
0$ as $r\rightarrow \infty$. Based on above analysis, one obtains
the following boundary condition for the function $\sigma(r)$ as
$r\rightarrow \infty$:
\begin{eqnarray}
\sigma &=& \sigma_v(\beta,\mu), \qquad \mathrm{for}\qquad T\leq T_{C} \qquad \mathrm{or}\qquad \mu\leq \mu_{C}, \label{condition3}\\
\sigma &=& \sigma_0, \qquad \mathrm{for}\qquad T_{C}<T \qquad
\mathrm{or}\qquad \mu_{C}< \mu.\label{condition4}
\end{eqnarray}
In order to investigate the behaviors of soliton solutions at finite
temperature and density, the set of coupled nonlinear differential
equations (\ref{equation1}), (\ref{equation2}) and
(\ref{equation31}) should be solved numerically with the boundary
conditions
Eqs.(\ref{condition1}),(\ref{condition2}),(\ref{condition3})  and
(\ref{condition4}). In the following discussion, we will focus on
two cases, one is that the temperature is $T\leq T_{C}$ as the
chemical potential is zero, the other is that the chemical potential
is $\mu\leq \mu_{C}$ by fixing the temperature at $T=50
\mathrm{MeV}$.

\begin{figure}
\includegraphics[scale=0.36]{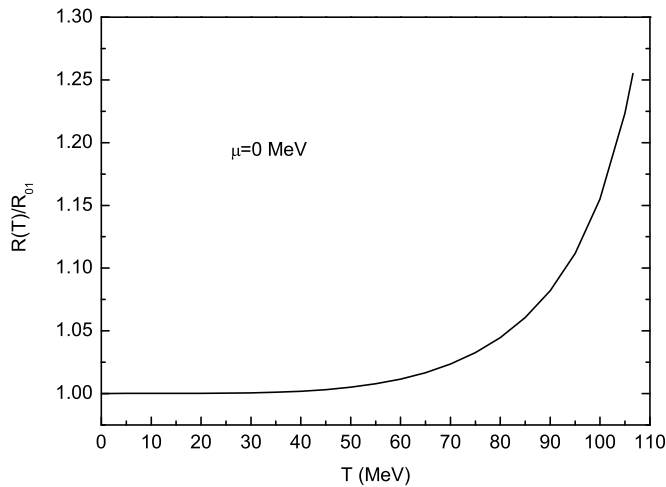}
\caption{\label{Fig:Fig8} Radius of the soliton bag as functions of
temperature when $T \leq T_{C}$ and $\mu=0 \mathrm{MeV}$. $R_{01}$
is the radius of the soliton bag for $T=0 \mathrm{MeV}$ and $\mu=0
\mathrm{MeV}$.}
\end{figure}

For the first case, in Fig.\ref{Fig:Fig7}, we plot the soliton
solutions by taking the temperatures as $T=0 \mathrm{MeV}, 60
\mathrm{MeV}, 90 \mathrm{MeV}$ and $106.6 \mathrm{MeV}$. From
Fig.\ref{Fig:Fig7}, we can see that with the temperature increasing,
$\sigma_v(\beta)$ is nearly constant near $\sigma_v$, while
$\sigma(0)$ is changed dramatically, and there exist stable soliton
bags for $T \leq T_C$. Also it can be seen in Fig.\ref{Fig:Fig7} the
radius of a stable soliton bag increases as temperature increases.
When the temperature is $T_{C}<T$, in Fig.\ref{Fig:Fig11}, we plot
the solutions by taking the temperatures as $T=110 \mathrm{MeV}, 120
\mathrm{MeV}, 123 \mathrm{MeV}$ and $123 \mathrm{MeV}$. As the
temperature is higher than the critical temperature $T_{C}$, the bag
constant is zero, the physical vacuum becomes unstable, and the
perturbative vacuum state is stable, the boundary condition for the
function $\sigma$ is changed accordingly, then the shapes of
solutions are very different from that of $T\leq T_{C}$. With the
increase of temperature, the soliton solutions tend to disappear.
From Fig.\ref{Fig:Fig11} we can see that, unlike the conventional
soliton solutions, here the solutions have the damping oscillation,
and we can not find soliton solution anymore.

Following the discussions of Ref.\cite{Mao:2006zp}, the radius of a
stable soliton bag increases as temperature increases. From
Eq.(\ref{radius}), in Fig.\ref{Fig:Fig8} we plot the variation of
radius as functions of temperature, which reveals that the radius of
the soliton bag does increase with increasing temperature. Once the
temperature is larger than the critical temperature $T_C$, the
physical vacuum becomes unstable while the perturbative vacuum state
is stable. Moreover, just as above mentioned, there is no soliton
solution when the temperature crosses over the critical temperature,
this means that the stable soliton bag is going to melt away and
disappear. Therefore there is no definite radius for $T > T_{C}$.

\begin{figure}
\includegraphics[scale=0.36]{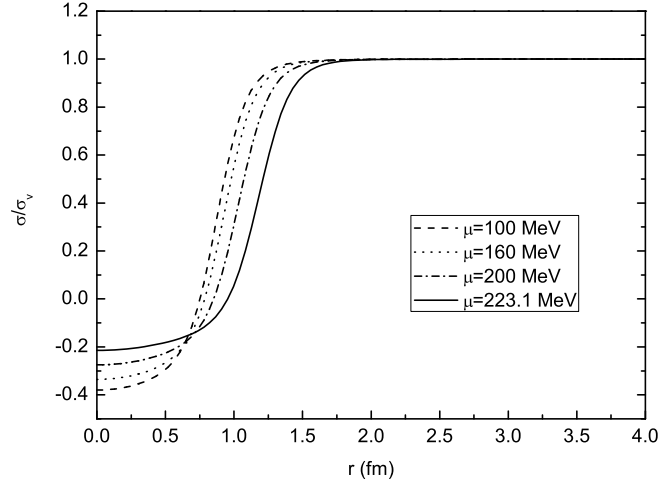}
\caption{\label{Fig:Fig9} The solutions for the set of coupled
nonlinear differential equations (\ref{equation1}),
(\ref{equation2}) and (\ref{equation31}) for different chemical
potential when $\mu \leq \mu_{C}$ and $T=50 \mathrm{MeV}$.}
\end{figure}

\begin{figure}
\includegraphics[scale=0.36]{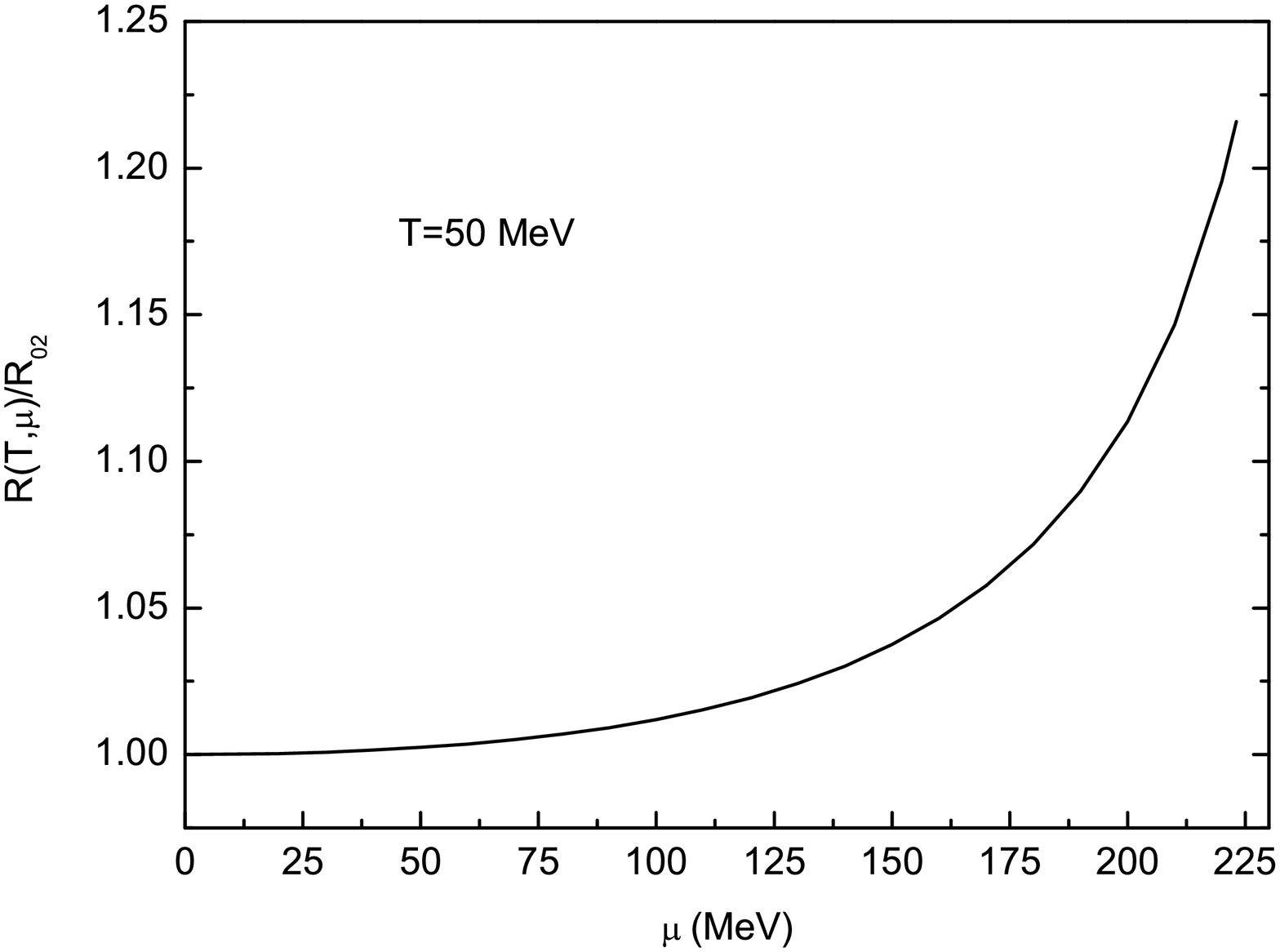}
\caption{\label{Fig:Fig10} Radius of the soliton bag as functions of
chemical potential when $\mu \leq \mu_{C}$ and $T=50 \mathrm{MeV}$.
$R_{02}$ is the radius of the soliton bag for $T=50 \mathrm{MeV}$
and $\mu=0 \mathrm{MeV}$.}
\end{figure}

\begin{figure}
\includegraphics[scale=0.36]{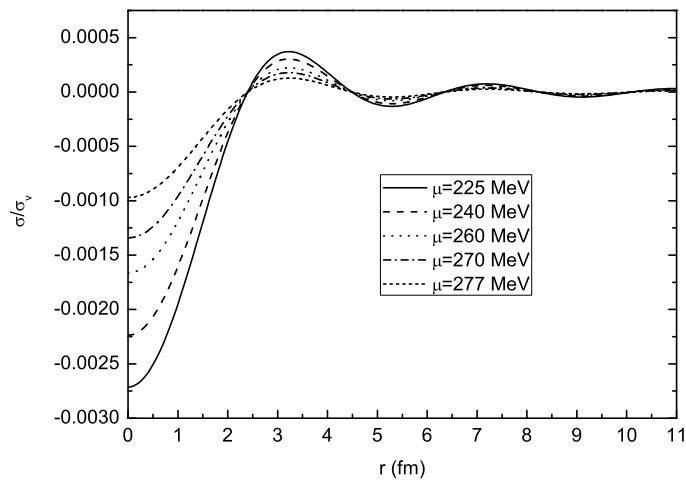}
\caption{\label{Fig:Fig12} The solutions for the set of coupled
nonlinear differential equations (\ref{equation1}),
(\ref{equation2}) and (\ref{equation31}) for different chemical
potential when $\mu>\mu_{C}$ by fixing the temperature $T=50
\mathrm{MeV}$.}
\end{figure}

Similar discussion can be applied to the second case, when the
temperature is fixed, the soliton solutions at different chemical
potentials are illustrated in Fig.\ref{Fig:Fig9}, where the chemical
potentials are taken as $\mu=100 \mathrm{MeV}$, $160 \mathrm{MeV}$
and $223.1 \mathrm{MeV}$ for a fixed temperature $T=50
\mathrm{MeV}$. With the chemical potential increasing, the radius of
a stable soliton bag increases too, we plot the radius of the
soliton bag versus chemical potential at a fixing temperature in
Fig.\ref{Fig:Fig10}. We can see clearly that the radius of the
soliton bag increases continuously with the increase of chemical
potential. When the chemical potential is larger than the critical
chemical potential $\mu_C$, due to the fact that the physical vacuum
becomes unstable and the perturbative vacuum state is stable, there
is no definite radius and the bag does not exist anymore. In order
to show that there is no stable soliton bag when the chemical
potential is above the critical chemical potential, the set of
coupled nonlinear differential equations (\ref{equation1}),
(\ref{equation2}) and (\ref{equation31}) should be solved
numerically with the boundary conditions
Eqs.(\ref{condition1}),(\ref{condition2}) and (\ref{condition4}) for
$\mu_C<\mu$ by fixing the temperature at $T=50 \mathrm{MeV}$. We
plot the solutions when the chemical potential is larger than the
critical chemical potential by taking the chemical potential as
$\mu=225 \mathrm{MeV}, 240 \mathrm{MeV}, 260 \mathrm{MeV}, 270
\mathrm{MeV}$ and $277 \mathrm{MeV}$ in Fig.\ref{Fig:Fig12}. When
the chemical potential is higher than the critical chemical
potential $\mu_{C}$, the bag constant is zero, the physical vacuum
becomes unstable, and the perturbative vacuum state is stable,
accordingly the boundary condition for the function $\sigma$ is
changed, then the shapes of solutions are very different from that
of $\mu\leq \mu_{C}$. With the increase of chemical potential, the
soliton solutions tend to disappear. From Fig.\ref{Fig:Fig12} it can
be seen that the solutions have the damping oscillation and there is
no soliton solution.

As mentioned in Sect.III,  $B(T)$ is defined as the difference
between the vacua inside and outside the soliton bag. For $T\leq
T_{C}$ and $\mu=0 \mathrm{MeV}$, the vacuum inside the soliton bag
is the metastable vacuum, the vacuum outside the soliton bag is the
real physical vacuum. This can be seen in Fig.\ref{Fig:Fig7}. As the
temperature is above the critical temperature $T_{C}$, the
metastable vacuum becomes the absolute one.  The behavior of
$\sigma(r)$ changes dramatically when $T>T_{C}$, i.e. $\sigma(r)$ is
very close to zero at any $r$, this can be seen from
Fig.\ref{Fig:Fig11}. Therefore, the values of the effective
potential are always close to their stable minimum at $\sigma=0$.
The state at $\sigma \sim \sigma_v(\beta)$ can no longer be realized
in this case. This means that the bag constant $B=0$. Based on above
analysis, there is no soliton solution and the bag constant $B(T)$
is zero for $T>T_{C}$, so there exists no more mechanism in
Friedberg-Lee model to confine the quarks, and the quarks are to be
deconfined when $T>T_C$. Accordingly, for $\mu\leq \mu_{C}$ and
$T=50 \mathrm{MeV}$, the vacuum inside the soliton bag is the
metastable vacuum, the vacuum outside the soliton bag is the real
physical vacuum shown in Fig.\ref{Fig:Fig9}. As the chemical
potential is above the critical chemical potential $\mu_{C}$, the
metastable vacuum becomes the absolute one. It can be seen from
Fig.\ref{Fig:Fig12}, the behavior of $\sigma(r)$ changes
dramatically when $\mu>\mu_{C}$, i.e. $\sigma(r)$ is very close to
zero at any $r$. Therefore, the values of the effective potential
are always close to its stable minimum at $\sigma=0$. The state at
$\sigma \sim \sigma_v(\beta)$ can not exist in this case, and the
bag constant $B=0$, hence there is no soliton solution and the bag
constant $B(T)$ is zero for $\mu>\mu_{C}$, no more mechanism in
soliton bag model can be applied to confine the quarks, then the
quarks are to be decofined in high chemical potential.

\section{Summary and discussion}

In this paper, we have studied the deconfinement phase transtion of
the Friedberg-Lee model at finite temperature and chemical
potential. Our studies are mainly concentrated on the effective
potential of the Friedberg-Lee model and the soliton solutions at
different temperatures and chemical potentials.We have gotten the
bag constant functions $B(T)$ and $B(T,\mu)$ for finite temperature
and density respectively. It is shown that the two minima of the
potential at zero temperature and chemical potential will be equal
at a certain temperature $T\simeq 106.6 \mathrm{MeV}$ or a certain
chemical potential $\mu \simeq 223.1 \mathrm{MeV}$, which are to be
defined as the critical temperature $T_C$ and the critical chemical
potential $\mu_C$ accordingly.

When the temperature $T>T_{C}$, the original perturbative vacuum
state becomes stable and the original physical vacuum state becomes
a metastable one. Because $\sigma(r)$ is very close to zero at any
$r$ for $T>T_{C}$, the effective potential always takes its value at
the absolute minimum $V(\sigma \sim 0)$. This gives $B(T)=0$ for
$T>T_{C}$. As the temperature approaches another higher temperature
$T \simeq 129 \mathrm{MeV}$, the effective potential has a unique
minimum. Similarly, for $\mu>\mu_{C}$, the original perturbative
vacuum state becomes stable and the original physical vacuum state
becomes a metastable one. Due to the fact that $\sigma(r)$ is very
close to zero at any $r$ for $\mu>\mu_{C}$, the effective potential
always takes its value at the absolute minimum $V(\sigma \sim 0)$.
This makes $B(T,\mu)=0$ for $\mu>\mu_{C}$. As the chemical potential
approaches another higher chemical potential $\mu \simeq 354
\mathrm{MeV}$, the effective potential has a unique minimum. Our
results are qualitatively similar to that obtained in conventional
soliton bag model at finite temperature\cite{Gao:1992zd,Mao:2006zp,
Li:1987wb, Wang:1989ex}.

In Friedberg-Lee model, the confinement of the quark requires the
existence of the soliton solution, and the soliton solution depends
on the effective potential at finite temperature and density. In
order to investigate the behavior of the soliton solution at finite
temperature and chemical potential, we numerically solve the set of
coupled nonlinear differential equations. Our results show that when
$T\leq T_{C}$ (or $\mu\leq \mu_C$), there exist the stable soliton
solutions in the model, however, when the temperature is higher than
the critical temperature $T_{C}\simeq 106.6 \mathrm{MeV}$ (or the
critical chemical potential $\mu_C \simeq 223.1 \mathrm{MeV}$),
there is only damping oscillation solutions and no soliton-like
solution exists, and the quark can not be confined by such solutions
anymore, then the confinement of quarks are removed and the
deconfined phase transition takes place. We also obtain that the
radius of the soliton bag does increase when the temperature or
density increases. At $T=T_C$ (or $\mu=\mu_C$) the soliton bag
disappears.

Since the soliton bag model has four adjustable parameters, the
numerical results of the critical temperature and the critical
chemical potential of the deconfinement phase are
parameter-dependent. Moreover in this work, we have only studied the
single nucleon properties at finite temperature and chemical
potential and have not considered the interactions of the nucleons.
In order to obtain a reasonable description of nuclear
interactions\cite{Guichon:1987jp}\cite{Saito:1994ki}\cite{Saito:2005rv},we
should consider the modifications of the Friedberg-Lee model to
include explicit meson degrees of freedom coupled linearly to the
quarks, which have been studied recently in Ref.\cite{Barnea:2000nu}
at zero temperature, it is of interest to extend their work to
finite temperature and density, and discuss the soliton solutions at
different temperatures and chemical potentials. All these works are
in progress.

\begin{acknowledgments}
The authors wish to thank Ru-Keng Su, Chen Wu, Pengfei Zhuang and
especially D.H. Rischke for valuable discussions. This work is
supported in part by the National Natural Science Foundation of
People's Republic of China (No.10747121) and the Department of
Education of Zhejiang Province under Grant No. 20070405.
\end{acknowledgments}

\end{document}